\documentclass[twocolumn,amsmath,aps]{revtex4}
\usepackage{graphicx}

\newcommand{\nl}{\nonumber \\}

\newcommand{\be}{\begin{equation}}
\newcommand{\ee}{\end{equation}}
\newcommand{\bea}{\begin{eqnarray}}
\newcommand{\eea}{\end{eqnarray}}

\newcommand{\Eq}[1]{Eq.\,(\ref{#1})}
\newcommand{\Eqs}[1]{Eqs.\,(\ref{#1})}

\newcommand{\la}{\langle}
\newcommand{\ra}{\rangle}
\newcommand{\dg}{\dagger}

\newcommand{\ep}{\epsilon}

\newcommand{\ti}{\tilde}
\newcommand{\mb}{\mbox}

\begin{document}
\draft

\title{ Quantum master equation scheme of time-dependent density functional theory
        to time-dependent transport in nano-electronic devices }

\author{Xin-Qi Li}
\email{xqli@red.semi.ac.cn}
\affiliation{State Key Laboratory for Superlattices and Microstructures,
         Institute of Semiconductors,
         Chinese Academy of Sciences, P.O.~Box 912, Beijing 100083, China}
\author{YiJing Yan}
\email{yan@ust.hk}
\affiliation{Department of Chemistry, Hong Kong University of
             Science and Technology, Kowloon, Hong Kong}
\date{\today}

\begin{abstract}
In this work a practical scheme is developed for the first-principles study
of time-dependent quantum transport.
The basic idea is to combine the transport master-equation with the well-known
time-dependent density functional theory.
The key ingredients of this paper include:
(i) the partitioning-free initial condition and the consideration of
the time-dependent bias voltages which base our treatment on the
Runge-Gross existence theorem;
(ii) the non-Markovian master equation for the reduced (many-body)
central system (i.e. the device);
and (iii) the construction of Kohn-Sham master equation for the
reduced single-particle density matrix, where a number of
auxiliary functions are introduced and their equations of motion
(EOM) are established based on the technique of spectral decomposition.
As a result, starting with a well-defined initial state, the
time-dependent transport current can be calculated simultaneously
along the propagation of the Kohn-Sham master equation and the EOM
of the auxiliary functions.\\
\\
PACS numbers: 72.10.Bg,72.90.+y
\end{abstract}
\maketitle

\section{ Introduction }

Quantum transport through nanostructures (e.g. the semiconductor quantum dots
or organic molecules) would play essential role in nano-electronic devices.
A rigorous treatment should handle the effect of the electronic structure
of the central {\it device}
as well as  the effect of the interface to the external contact.
This calls for combining the theory of quantum transport
with the first-principles calculation of electronic structures.
In recent years, considerable efforts have been focused on
the density-functional theory (DFT) based simulations on such transport
devices \cite{Lan00,Heu03,Wan03}.
In particular, two types of formalisms were involved:
one is the Lippmann-Schwinger formalism by Lang and coworkers \cite{Lan95};
the other is the first-principles non-equilibrium Green's function (nGF) technique
\cite{Tay01,Ke04,Den04,Bra02,Xue01}.
However, all of these studies were restricted to the steady-state transport under
time-independent bias voltages.

On the other hand, time-dependent transport phenomenon are of interest
in various contexts, such as the single electron pumps and turnstiles,
switching-on transient behaviors, and ac response in the applications
of high-frequency amplifiers or detectors.
Moreover, it will be desirable if one is able to model the
nano-electronic circuit elements with resistors, capacitors, and inductors.
To ascribe appropriate {\it quantum} capacitances and inductances to the nano-circuits,
the study of time-dependent quantum transport is needed.
With this motivation, pioneering work has been carried out
\cite{CT91,Lan91,Pas92,Iva9395,Win93-96,Sch9495},
where the time-dependence enters via self-consistent parameters
and the entire study was largely based on over-simplified model description.

Obviously it is desirable to extend this kind of study to
the level of {\it first-principles} simulation.
To this end, a necessary element to be combined is seemingly
the time-dependent density functional theory (TDDFT) \cite{RG84},
which is a generalization of the well-known
(ground state) static DFT \cite{HK64,KS65}.
Similar to the static DFT, the fundamental variable in TDDFT
is no longer the many-body wave function but the density.
This time-dependent density is determined by solving an auxiliary set of
{\it noninteracting} time-dependent Schr\"odinger equations,
say, the Kohn-Sham equations.
The nontrivial part of the many-body interaction is contained in the so-called
exchange-correlation (xc) potential, for which reasonably good approximations exist.
Since the foundation of TDDFT,
an enormous amount of progress has been made and the theory was applied to
a large number of problems in physics and chemistry.
In particular, even the simplest approximation to the xc potential,
i.e., the so-called adiabatic LDA, can yield remarkably good results.
Nevertheless, despite the wide range of applications, the TDDFT has been mostly
limited to isolated systems.
To our knowledge, its application to quantum open systems
(e.g. in quantum transport) just begins very recently.
These interesting efforts include either employing formally the nGF formalism
\cite{Ste04,Chen06}, or propagating wave-function at zero temperature \cite{Ste05},
or introducing electron-phonon scattering to balance the external field
in order to reach a stationary current \cite{Bur05,Geb05}.

In this work, by combining the TDDFT with our recently constructed quantum master
equation formalism to quantum transport \cite{Li05a,Li05b,Luo06,Li04},
we attempt to develop an alternative scheme.
The master equation approach to mesoscopic transport can be
dated back to the classical
rate equation \cite{Gla88} and its quantum generalization \cite{Gur96}.
More recent work based on the master equation approach can be
found, for instance, in Refs.\ \onlinecite{GLD0406,Bru03,Jau04,Wel05}.
This approach is of interest not only by its conceptual difference from the
conventional mesoscopic transport theory, e.g., the
Landauer-B\"uttiker theory or the non-equilibrium Green's function (nGF)
technique \cite{Dat95,Hau96}, but also due to its convenience in applications.
More importantly, since the master equation is time-dependent,
it seems thus a natural framework for the study of time-dependent transport.
This is in fact the main motivation for us to combine it with the TDDFT
to construct a time-dependent transport scheme at the level of first-principles.

In the standard TDDFT,
the propagation of the system state is described by the time-dependent Kohn-Sham equation,
whose solution is used to calculate the (time-dependent) density and the
effective Hamiltonian of the non-interacting Kohn-Sham system.
In the context of quantum transport,
using the same idea of TDDFT, we first map the {\it entire system}
(i.e. the central device plus the electrodes) to the non-interacting Kohn-Sham system,
then trace out the degrees of freedom of the electrodes.
As a consequence, directly related to the time-dependent electron density,
we obtain a master equation for the reduced density matrix of the device,
which is a counterpart of the well known Kohn-Sham Schr\"odinger equation.
We notice that this treatment can be {\it exact} in principle.
That is, the entire system of the {\it device plus electrodes} evolves under
the influence of bias voltage, starting from a well-defined initial state.
Then, the requirement of the Runge-Gross {\it existence theorem} is satisfied.
Of course, as any other (time-dependent) transport theories,
proper approximate consideration is needed for the biased electrodes,
which enables us to eliminate the degrees of freedom of electrodes
and obtain a master equation for the reduced state of the central device.

The remainder of the paper is organized as follows.
In next section we first specify the transport setup and some physical considerations,
then outline the main result of the master equation approach to transport.
In this work, instead of the Markovian prescription \cite{Li04,Li05a,Li05b,Luo06},
we would like to adopt the scheme of non-Markovian master equation
for reasons to be specified later in the main text.
For completeness, a brief derivation for the non-Markovian ``$n$"-resolved
master equation is provided in Appendix A.
In Sec.\ III we combine the TDDFT scheme with the multi-particle-state
master equation obtained in Sec.\ II.
By introducing the single-particle density matrix and a number of
auxiliary functions associated with the spectral decomposition,
we establish the central result of this work, say, the Kohn-Sham master equation
for the reduced single-particle density matrix, and the propagating equations
for the auxiliary functions. These quantities would suffice the calculation
of the transport current.
Also, a brief description for the technique of spectral decomposition will
be presented in Appendix B.
In Sec.\ IV we describe how the effect of inelastic electron-phonon scattering
in the device can be conveniently included into the Kohn-Sham master equation.
Finally, summary and discussions are given in Sec.\ V.


\section{ Transport Master Equation }

In this section we shall first present the transport model
together with some physical considerations on it,
then outline the main result of the transport master equation
in many-particle-state Hilbert space, and put its derivation
in Appendix A.
All of these will form the basis of the TDDFT scheme to quantum transport
of next section.

\subsection{Model Consideration}

Conventionally, mesoscopic transport setup can be described by the transfer Hamiltonian
\bea\label{H-ms}
H &=& H_C(a_{\mu}^{\dg},a_{\mu})+\sum_{\alpha=L,R}\sum_{\mu k}
       \epsilon_{\alpha\mu k}d^{\dg}_{\alpha\mu k}d_{\alpha\mu k}  \nl
  & &  + \sum_{\alpha=L,R}\sum_{\mu k}(t_{\alpha\mu k}a^{\dg}_{\mu}
       d_{\alpha\mu k}+\rm{H.c.}) .
\eea
Here $H_C$ is the {\it device} Hamiltonian of the central region,
which can be rather general, e.g., including electron-electron and
electron-phonon interactions,
and may be extended to contain a few interface atoms of electrodes
in the context of such as molecular transport devices.
The second and third terms describe, respectively, the (left and right) electrodes
and the tunneling between them and the central device.
Note that in the above Hamiltonian, as widely adopted in transport studies,
the electrode electrons are treated as noninteracting after absorbing
the interactions into self-consistent potential.
This noninteracting feature of the electrodes
is in fact the basis of the Landauer-type theory,
which is the cornerstone of quantum transport.
Also, the electrode electrons are described in Bloch-state representation.
This choice is largely for the convenience of formulation.
To make contact with the first-principles calculation,
the local Wannier-state representation will be more appropriate.
As will be shown in Sec.\ III, the conversion between them can be easily
implemented via the link of self-energy matrix.


Since our major concern is the time-dependent and transient behaviors,
care should be made for the choice of {\it initial condition}.
In practice, two types of initial states were adopted:
(i)
In the nGF approach to quantum transport, Caroli {\it et al.} \cite{Car71}
considered the two leads as isolated subsystems with different chemical potentials
in the remote past.
The current starts to flow through the system once the contacts between the device
and the leads have been established.
This treatment of partitioning is seemingly a bit {\it fictitious},
since in a real experiment the whole system is in thermodynamic equilibrium
before an external bias is applied deep inside the electrodes.
Later on this scheme was adopted by Meir and Wingreen {\it et al.} to
obtain the steady-state current through an interacting central device \cite{Mei92},
and also to time-dependent phenomena \cite{Win93-98}.
(ii)
Conceptually differing from the one by Caroli {\it et al.},
an alternative scheme was suggested by Cini \cite{Cin80}, in which the central
device is contacted and in thermodynamic equilibrium before an external
time-dependent disturbance (i.e. the bias voltage) is switched on.
This type of initial condition was recently applied by Stefanucci {\it et al}
\cite{Ste04a,Ste04,Ste05}.

In this work, we involve the Cini's initial condition as follows.
Initially, before the bias voltage is switched on,
we let the central device be contacted with the leads and in
thermodynamic equilibrium.
As the bias is switched on, the external potential and the disturbance
caused by the device region are {\it screened} deep inside the electrodes,
thus the density inside the electrodes approaches the equilibrium bulk-one.
This leads to an enormous simplification since the initial-state
self-consistency is not disturbed far away from the device.
This picture holds to be valid
provided that the driving frequency is smaller than the plasma frequency,
which is tens of THz in typical doped semiconductors.
The bias is established entirely across the device by charge accumulation
and depletion near the electrode-device interfaces.
The formation of these charge layers then causes a rigid shift of the conduction band
of the electrodes, but remains the population unchanged as the initial one.
As an equivalent and more convenient description, the above consideration
is typically replaced by the statement
that the reservoir electrons are {\it always} in local thermal equilibrium
characterized by the chemical potentials $\mu_{L/R}(t)$ which define
the bias voltage $V(t)$ via $eV(t)=\mu_L(t)-\mu_R(t)$.
In this way, the initial equilibrium state is violated, and the system starts evolving
under the driving of the bias voltage, leading to a time-dependent transport current.
In practice, there exist two types of bias voltages:
One is the slow time-dependent modulation of the bias voltage, which is
appropriate for the study of ac response;
another is the one {\it suddenly} switched-on, i.e.,
$eV(t)\simeq (\mu_L-\mu_R)\Theta(t)$,
which corresponds to the initial setup for studying the transient behaviors.

Notice that in the above treatment,
we explicitly keep the chemical potentials (Fermi levels) of the electrodes
at {\it instant values} determined by the time-dependent bias voltage,
and let the electrode reservoirs always in local thermal equilibrium.
Physically, this treatment properly take into account
the {\it closed nature} of the transport circuit
and the {\it rapid relaxation} in the electrode reservoirs.
In Cini's treatment, two spatially homogeneous and time-dependent
{\it electrical potentials} are added to the electrode Hamiltonians.
The use of the initial occupation in the electrodes then corresponds to
the {\it initial} rigid shift of the Fermi levels.
However, in the subsequent evolution
it seems unclear how the rapid relaxation in the electrode
reservoirs is included, and how the (approximate) local thermal
equilibrium of the reservoirs is guaranteed.

\subsection{``$n$"-Resolved Master Equation and Transport Current}

The key idea of the quantum master equation approach to transport
is to view the {\it device} as an open dissipative system, and
the electrodes as its environment.
Following the standard treatment of quantum dissipation theory,
we introduce the reservoir operators
 $F_{\mu} = \sum_{\alpha k} t_{\alpha\mu k}d_{\alpha\mu k}
  \equiv f_{L\mu} + f_{R\mu}$.
Accordingly, the tunneling Hamiltonian $H'$ reads
$ H' = \sum_{\mu} \left( a^{\dg}_{\mu} F_{\mu}
       + \rm{H.c.}\right) $.
Let us then consider the entire system evolution starting from such
{\it initial state} as discussed above, i.e., the central device
in a state formed under zero bias {\it in the presence
of device-electrode coupling}, and the electrodes in local thermal
equilibrium states defined by the {\it initial chemical potentials}.
In the later on evolution, we treat $H'$ as perturbation.
Note that this {\it does not} mean the partitioning
of the device from the electrodes {\it before the bias voltage is switched on}.

Treating $H'$ perturbatively under the second-order Born approximation
\cite{Li04,Yan98,Xu02},
and performing the ``$n$"-resolved electrode states average as shown in Appendix A,
we obtain
\bea\label{ME-1}
\dot{\rho}^{(n)}
   &=&  -i {\cal L}\rho^{(n)} - \sum_{\mu}
   \left\{ [a_{\mu}^{\dg}A_{\rho^{(n)}\mu}^{(-)}
   +A_{\rho^{(n)}\mu}^{(+)}a_{\mu}^{\dg} \right. \nl
   & &         - A_{\rho^{(n)}L\mu}^{(-)}a_{\mu}^{\dg}
            - a_{\mu}^{\dg}A_{\rho^{(n)}L\mu}^{(+)}        \nl
& & \left. - A_{\rho^{(n-1)}R\mu}^{(-)}a_{\mu}^{\dg}
   - a_{\mu}^{\dg}A_{\rho^{(n+1)}R\mu}^{(+)}]+{\rm H.c.} \right\} ,
\eea
where
$A_{\rho^{(n)}\mu}^{(\pm)}=\sum_{\alpha=L,R}A_{\rho^{(n)}\alpha\mu}^{(\pm)}$, with
\bea\label{A-rho-n-t}
A_{\rho^{(n)}\alpha\mu}^{(-)}(t) &=& \sum_{\nu}\int^{t}_{0} dt'
    C^{(-)}_{\alpha\mu\nu}(t,t'){\cal G}(t,t')[a_{\nu}\rho^{(n)}(t')]  \nl
A_{\rho^{(n)}\alpha\mu}^{(+)}(t) &=& \sum_{\nu}\int^{t}_{0} dt'
    C^{(+)}_{\alpha\nu\mu}(t',t){\cal G}(t,t')[\rho^{(n)}(t')a_{\nu}] . \nl
\eea
The bath correlation functions read
$C^{(-)}_{\alpha\mu\nu}(t,t')=\la f_{\alpha\mu}(t)f^{\dg}_{\alpha\nu}(t')\ra$,
and $C^{(+)}_{\alpha\nu\mu}(t',t)=\la f^{\dg}_{\alpha\nu}(t')f_{\alpha\mu}(t)\ra$.
Note that here, differing from Ref.\ \onlinecite{Li04},
we have adopted the non-Markovian (i.e. time non-local)
form of master equation.

With the knowledge of $\rho^{(n)}(t)$, one is readily able to compute
the various transport properties, such as the transport current and
noise spectrum \cite{Li05a,Li05b,Luo06,Li04}.
In this work, we focus on the calculation of transport current.
Based on the probability distribution function
$P(n,t)\equiv\rm{Tr}[\rho^{(n)}(t)]$,
inserting \Eq{ME-1} into $I(t)=e\sum_n n\dot{P}(n,t)$
gives rise to
\bea\label{I-t}
I(t) = 2e \sum_{\mu} \mb{Re} \mb{Tr} \left[ a^{\dg}_{\mu}
 \left( A^{(-)}_{\rho R\mu}(t)-A^{(+)}_{\rho R\mu}(t)\right)\right] .
\eea
Here, $A^{(\pm)}_{\rho R\mu}(t)$ are the same as defined by \Eq{A-rho-n-t}
with the replacement of $\rho^{(n)}$ by
the unconditional density matrix $\rho=\sum_n \rho^{(n)}$.
Summing up \Eq{ME-1} over ``$n$", we obtain
\bea\label{rho-t}
 \dot{\rho} = -i {\cal L}\rho - \sum_{\mu}
        \left\{ [a_{\mu}^{\dg},A_{\rho\mu}^{(-)}(t)
        -A_{\rho\mu}^{(+)}(t)]+ \rm{H.c.} \right\}  ,
\eea
where $A^{(\pm)}_{\rho \mu}(t)=\sum_{\alpha=L,R} A^{(\pm)}_{\rho \alpha\mu}(t)$.
In principle, to carry out $A^{(\pm)}_{\rho\alpha \mu}(t)$ in this non-Markovian scheme,
one should know $\rho(t')$ for all the time $t'\in[0,t)$,
which is in fact the non-Markovian nature (i.e. the time non-locality or memory effect).
However, as will be shown in the following,
based on the technique of spectral decomposition,
we are going to establish the EOM of $A^{(\pm)}_{\rho \alpha\mu}(t)$.
Then, the combined propagation of $A^{(\pm)}_{\rho \alpha\mu}(t)$ and $\rho(t)$
will give the full time-dependent solution for the transport problem under study.
Note that these coupled equations will be time local, their propagation is thus quite
straightforward.
In contrast, as shown in Ref.\ \onlinecite{Li05b},
if we start with a Markovian scheme for the many-particle-state master equation,
the {\it time-local} scheme for the {\it reduced single-particle density matrices}
is seemingly impossible to be constructed, and
the time nonlocal feature will make the practical propagation quite difficult.


\section{TDDFT Scheme}

The transport master equation constructed in the previous section
is defined in many-particle Hilbert space.
This will restrict its applicability to few-states systems,
because the huge dimensions of many-particle Hilbert space
for large-scale systems would make the problem intractable.
In this section, in the spirit of TDDFT we recast the many-particle
interacting system to a Kohn-Sham noninteracting system, and
establish the corresponding transport master equation, which is defined
in the single-particle-state Hilbert space with dimensions greatly reduced.

\subsection{General Consideration}

Taking the entire system of {\it electrodes plus device} into account
and starting with the well-defined initial state as described in Sec.\ II(A),
the Runge-Gross theorem implies a {\it one-to-one correspondence} between
the electron density and the potential function.
Then, the electron density distribution uniquely determines
all the (transport) properties of the system.
The significant role of the Runge-Gross {\it existence theorem}
is to guarantee the construction of a proper noninteracting Kohn-Sham system.

For the electrodes, the Kohn-Sham mapping to noninteracting system
is relatively simple.
In practice, the interactions in the electrodes are absorbed into a self-consistent
potential, and the electrons in the electrodes are treated as {\it noninteracting}.
As a matter of fact, the noninteracting feature of the electrodes
is the basis of the Landauer-type transport theory,
which is the cornerstone of quantum transport.

The part of the system that we are going to treat carefully is the (extended) device.
Its Kohn-Sham counterpart can be constructed as follows.
In principle, if we knew the many-particle density operator $\rho(t)$
of the device, we could
introduce the {\it reduced single-particle} (RSP) density matrix,
$\sigma_{\mu\nu}(t)\equiv\rm{Tr}[a^{\dg}_{\nu}a_{\mu}\rho(t)]$.
However, rather than solving $\rho(t)$, we should calculate
$\sigma_{\mu\nu}(t)$ directly from its equation of motion,
which is the central gaol of this work.
Within the TDDFT framework \cite{RG84}, we map the device {\it interacting}
Hamiltonian to the non-interacting Kohn-Sham one as
\bea\label{Fock}
h_{mn}(t) = h^{0}_{mn}(t)+v^{\rm xc}_{mn}(t)
      +\sum_{ij}\sigma_{ij}(t)V_{mnij}.
\eea
In first-principles calculation the state basis is usually
chosen as the local atomic orbitals, $\{ \phi_m({\bf r}), m=1,2,
\cdots \}$.
Here $h^0(t)$ is the non-interacting Hamiltonian which can be in
general time-dependent; $V_{mnij}$ is the two-electron Coulomb
integral, $V_{mnij}=\int d{\bf r}\int d{\bf r'}\phi^*_m({\bf
r})\phi_n({\bf r}) \frac{1}{|{\bf r}-{\bf r'}|}\phi^*_i({\bf
r'})\phi_j({\bf r'})$; and $v^{\rm xc}_{mn}(t)=\int d{\bf r}
\phi^*_m({\bf r})v^{\rm xc}[n]({\bf r},t)\phi_n({\bf r})$, with
$v^{\rm xc}[n]({\bf r},t)$ the exchange-correlation potential,
which is defined by the functional derivative of the the
exchange-correlation functional $A^{\rm xc}$. In practice,
especially in the time-dependent case, the unknown functional
$A^{\rm xc}$ can be approximated by the energy functional $E^{\rm
xc}$, obtained in the Kohn-Sham theory and further with the local
density approximation (LDA).
Note that the density function $n({\bf r},t)$ appearing in the
Kohn-Sham Hamiltonian is related to the RSP density matrix via
$n({\bf r},t)=\sum_{mn}\phi_m({\bf r})\sigma_{mn}(t)\phi^*_n({\bf r})$.
Thus, the entire Kohn-Sham system is described by the noninteracting
electrodes, the device Hamiltonian $H(t)=\sum_{mn}h_{mn}(t)a^{\dg}_ma_n$,
and the coupling between them.
Starting with this entire Kohn-Sham Hamiltonian and repeating the derivation
in Sec. II (B) will lead to the same formal result presented there.
The only difference is that now the Kohn-Sham Hamiltonian $H(t)$
is noninteracting, which enables us to establish the EOM for
$\sigma_{\mu\nu}(t)$.

\subsection{Kohn-Sham Master Equation}

In this subsection, for the Kohn-Sham system, we recast the master equation
\Eq{rho-t} into the EOM of $\sigma_{\mu\nu}(t)$ and a set of auxiliary functions,
and re-express the transport current in terms of them.
For brevity, we tentatively restrict our derivation to the special case of
{\it constant} bias voltage. Generalization to time-dependent voltages is
straightforward, and will be discussed in Sec.\ III (D).
For constant bias voltage, the correlation functions $C^{(-)}_{\alpha\mu\nu}(t,t')$
and $C^{(+)}_{\alpha\nu\mu}(t',t)$ only depend on the difference of times.
By employing the technique of {\it spectral decomposition} as described in Appendix B,
they can be formally decomposed in the sum of exponential functions
\bea \label{Ctt+-}
C^{(-)}_{\alpha\mu\nu}(t,t')
 &=& \sum_k \lambda^{(-)k}_{\alpha\mu\nu} e^{\gamma^{(-)k}_{\alpha\mu\nu}(t-t')} , \nl
C^{(+)}_{\alpha\nu\mu}(t',t)
 &=& \sum_k \lambda^{(+)k}_{\alpha\nu\mu} e^{\gamma^{(+)k}_{\alpha\nu\mu}(t-t')} ,
\eea
where the parameters $ \lambda^{(\pm)k}$ and $\gamma^{(\pm)k}$
are uniquely determined by the spectral decomposition.
Furthermore, in addition to  $\sigma_{\nu\mu}(t)$, we introduce
\begin{subequations}
\bea
\ti{\sigma}_{\nu\mu}(t,t')
&=& {\rm Tr}\{ a^{\dg}_{\mu} {\cal G}(t,t')
    [a_{\nu}\rho(t') ]\}  ,  \\
\bar{\sigma}_{\nu\mu}(t,t')
&=& {\rm Tr}\{ a^{\dg}_{\mu} {\cal G}(t,t')
    [\rho(t')a_{\nu} ]\} ,
\eea
\end{subequations}
and the auxiliary functions
\begin{subequations}
\bea
A^{(-)k}_{\alpha\mu\nu\mu'}(t) &=& \int^{t}_{0} dt'
    \lambda^{(-)k}_{\alpha\mu\nu} e^{\gamma^{(-)k}_{\alpha\mu\nu}(t-t')}
    \ti{\sigma}_{\nu\mu'}(t,t')  ,  \\
A^{(+)k}_{\alpha\mu\nu\mu'}(t) &=& \int^{t}_{0} dt'
    \lambda^{(+)k}_{\alpha\nu\mu} e^{\gamma^{(+)k}_{\alpha\nu\mu}(t-t')}
     \bar{\sigma}_{\nu\mu'}(t,t') .
\eea
\end{subequations}
Simple algebra leads to the EOM of $\ti{\sigma}(t,t')$ and $\bar{\sigma}(t,t')$:
\begin{subequations}
\bea
\partial_{t} \ti{\sigma}_{\nu\mu}(t,t')
&=& i \left[\ti{\sigma}(t,t') h(t)\right]_{\nu\mu} ,  \\
\partial_{t} \bar{\sigma}_{\nu\mu}(t,t')
&=& i \left[\bar{\sigma}(t,t') h(t)\right]_{\nu\mu} .
\eea
\end{subequations}
Accordingly,
the EOM of $A^{(\pm)k}_{\alpha\mu\nu\mu'}(t)$ can be straightforwardly obtained:
\begin{subequations}\label{A+-kt}
\bea
& & \partial_t A^{(-)k}_{\alpha\mu\nu\mu'}(t)
  =  \lambda^{(-)k}_{\alpha\mu\nu} \sigma_{\nu\mu'}(t)    \nl
& & + \gamma^{(-)k}_{\alpha\mu\nu} A^{(-)k}_{\alpha\mu\nu\mu'}(t)
   + i \sum_m A^{(-)k}_{\alpha\mu\nu m}(t) h_{m\mu'}(t) ,  \\
& & \partial_t A^{(+)k}_{\alpha\mu\nu\mu'}(t)
  =  \lambda^{(+)k}_{\alpha\nu\mu} [\delta_{\nu\mu'}-\sigma_{\nu\mu'}(t)]    \nl
& & + \gamma^{(+)k}_{\alpha\nu\mu} A^{(+)k}_{\alpha\mu\nu\mu'}(t)
   + i \sum_m A^{(+)k}_{\alpha\mu\nu m}(t) h_{m\mu'}(t) .
\eea
\end{subequations}
Note that the RSP density matrix $\sigma_{\nu\mu}(t)$ appears in the EOM of
these auxiliary functions.
To close the EOM for the transport problem,
we need to derive the EOM of $\sigma_{\nu\mu}(t)$.
Using the identity
$\rm{Tr} \{ a^{\dg}_{\mu}a_{\nu} \sum_{\mu'}[a^{\dg}_{\mu'},A_{\mu'}] \}
 = \rm{Tr} [a^{\dg}_{\mu}A_{\nu}] $,
we easily obtain
\bea\label{sgm-t}
\dot{\sigma}(t) = -i [h(t),\sigma(t)]
         -\sum_{\alpha} \left[ M_{\alpha}(t)+ \rm{H.c.}\right] ,
\eea
where the matrix $M_{\alpha}(t)$ is defined via its elements as\bea
[M_{\alpha}]_{\nu\mu}(t) = \sum_{k\nu'} \left[ A^{(-)k}_{\alpha\nu\nu'\mu}(t)
          - A^{(+)k}_{\alpha\nu\nu'\mu}(t)   \right] . \nonumber
\eea
Propagating \Eqs{A+-kt} and (\ref{sgm-t}) is numerically straightforward,
and the most difficult task arising from the memory kernel in usual
non-Markovian dynamics is avoided.
Finally, the transport current is simply related to the auxiliary functions as
\bea
I(t) = 2e \sum_{k\mu\nu} {\rm Re} \left[ A^{(-)k}_{R\mu\nu\mu}(t)
       - A^{(+)k}_{R\mu\nu\mu}(t)  \right] ,
\eea
which is output {\it automatically} along the time propagation of the above EOM.

\subsection{Spectral Density Function}

The treatment in the previous subsection is based on the {\it spectral decomposition} of
$\Gamma^{\alpha}_{\mu\nu}(\epsilon)=2\pi\sum_k
       t^*_{\alpha\mu k}t_{\alpha\nu k}\delta(\ep-\ep_k)$, as shown in Appendix B.
Since the spectral decomposition is rooted in a technique of numerical fit,
it is thus in principle suited for arbitrary shape of the
spectral density function $\Gamma^{\alpha}_{\mu\nu}(\epsilon)$. In particular,
it is not limited by the conventional wide-band approximation.
This advantage makes us be able to incorporate the electronic structure of the electrodes
which is obtained from other sophisticated methods.
One possible way of calculating $\Gamma^{\alpha}_{\mu\nu}(\epsilon)$
is based on a semi-empirical tight-binding model for the electrodes,
and using the surface Green's function technique.
Some details of this method can be found in Ref.\ \onlinecite{Dat03}.
Here we would like only to outline the key idea for completeness.

Formally, in matrix form, the spectral density function is related with
the {\it self-energy matrix} $\Sigma^{\alpha}$ via
\bea
\Gamma^{\alpha} = i (\Sigma^{\alpha} -\Sigma^{\alpha\dg}) .
\eea
Further, the self-energy matrix can be calculated by
\bea
\Sigma = t g t^{\dg} ,
\eea
where ``$t$" is the coupling matrix between the ``edge atoms" of the electrode
and the (extended) device,
and $g$ is the {\it surface Green's function} of the electrode.
Recursively, $g$ can be obtained via the following Dyson equation
\bea
g^{-1}=g_0^{-1}-\ti{\Sigma}= g_0^{-1}-\ti{t} g \ti{t}^{\dg} ,
\eea
where ``$\ti{t}$" is the coupling matrix between nearest-neighbor layers
in the electrode.

In practice, care should be paid to the dimensions of the matrices, i.e.,
there are two types of orbital labels:
one is restricted to the device edge-orbitals;
and another is over all the device orbitals.
From the definition of each matrix, the size of its every dimension
can be identified accordingly.

\subsection{Time-Dependent Voltage}

The derivation in Sec.\ III (B) corresponds to time-independent bias voltage.
That is, after a constant voltage is switched on, the chemical potentials
in electrodes remain unchanged. This implies that the correlation functions
$C_{\alpha\mu\nu}^{(\pm)}(t,t')$ are of time-translational invariance
during the later evolution. The description under this assumption can be
applied to the study of {\it transient} behaviors \cite{Hau96}.

Another important time-dependent setup is applying modification
on the bias voltage, such as in the study of {\it ac response}.
As have discussed previously, under proper conditions,
the effect of time-dependent voltage can be approximately described by
{\it rigid shifts} of the conduction bands of the electrodes,
i.e., $\epsilon_{k\alpha}(t)=\epsilon_{k\alpha}+\Delta_{\alpha}(t)$,
and keeping the occupation on each state unchanged.
Accordingly, the correlation functions read
\begin{equation}\label{CCtt'}
\left[ \begin{array}{c}
C_{\alpha\mu\nu}^{(-)}(t,t') \\
C_{\alpha\nu\mu}^{(+)}(t',t) \\ \end{array} \right]
= \left[ \begin{array}{c}
\bar{C}_{\alpha\mu\nu}^{(-)}(t-t') \\
\bar{C}_{\alpha\nu\mu}^{(+)}(t-t') \\ \end{array} \right]
e^{-i\int^{t}_{t'}dt_1\Delta_{\alpha}(t_1)} ,
\end{equation}
where $\bar{C}_{\alpha\mu\nu}^{(\pm)}(t-t')$ are the counterparts
in the absence of time-dependent voltage shift (i.e. $\Delta_{\alpha}(t)=0$),
and can be obtained by using the surface Green's function technique
as described in the previous subsection.

Replacing the correlation functions in Sec.\ III (B) with \Eq{CCtt'},
all the equations obtained there can be formally re-derived,
except with only a minor modification on the second terms of the l.h.s. of
\Eq{A+-kt}, i.e.,
$\gamma^{(\pm)k}_{\alpha\mu\nu}\rightarrow
\gamma^{(\pm)k}_{\alpha\mu\nu}-i\Delta_{\alpha}(t)$.
The advantage of the proposed TDDFT scheme in Sec.\ III (B) is thus prominent,
since in other conventional treatments the time-dependent
transport is usually regarded more difficult than its stationary counterpart,
owing to the lack of time-translational invariance.
However, in our scheme, {\it no extra efforts} are needed
for time-dependent voltage in propagating
the EOM of the RSP density matix $\sigma_{\mu\nu}(t)$ and the auxiliary functions
$A^{(\pm)k}_{\alpha\mu\nu}(t)$.

\section{Electron-Phonon Interaction}

In this section we consider the issue of {\it inelastic scattering}
in the device. Still within the TDDFT framework,
we assume the Kohn-Sham subsystem of the device
being coupled to a phonon bath ($H_{\rm ph}=\sum_{q}\omega_qb_q^{\dg}b_q$).
In general, the electron phonon interaction Hamiltonian has the form
\bea
H_{\rm e-ph}&=&\sum_q\sum_{mn}\gamma_{qmn}(b_q^{\dg}+b_q)a^{\dg}_ma_n \nl
&\equiv& \sum_{q\kappa}(W_{\kappa}f^{\dg}_{q\kappa}
         +W^{\dg}_{\kappa}f_{q\kappa}),
\eea
where for brevity we introduce $W_{\kappa}\equiv W_{mn}=a^{\dg}_m a_n$,
and $f_{q\kappa}$ is defined accordingly.
Here and in the remainder of this section,
we adopt the electronic {\it eigenstate basis} of the device Kohn-Sham
Hamiltonian {\it before the bias voltage is applied}.
This choice has the advantage of making the electronic state transition
due to phonon scattering clearly defined.

Formally, the {\it many-electron-state} master equation for the device
can be expressed as
\bea
\dot{\rho}=-i{\cal L}\rho-{\cal R}_{\rm e}\rho-{\cal R}_{\rm ph}\rho .
\eea
In this equation, the term ${\cal R}_{\rm e}\rho$
describes the electrode effect on the device,
and ${\cal R}_{\rm ph}\rho$ stems from the effect of electron-phonon interaction.
In the previous sections, we have focused on the term ${\cal R}_{\rm e}\rho$,
by performing a non-Markovian treatment at the level of second-order Born approximation.
For ${\cal R}_{\rm ph}\rho$, one can in principle perform the same treatment,
by also applying the spectral decomposition technique for the phonon bath.
Nevertheless, for simplicity we would like to treat the electron-phonon interaction
under the Markovian approximation as usual.
Following Ref.\ \onlinecite{Xu02}, we have
\bea\label{Rrho-ph}
{\cal R}_{\rm ph}\rho
= \frac{1}{2}\sum_{\kappa}\left\{
  [W^{\dg}_{\kappa},\ti{W}^{(-)}_{\kappa}\rho
  -\rho\ti{W}^{(+)}_{\kappa} ]+{\rm H.c.}\right\} .
\eea
At the transition energy $\omega_{\kappa}\equiv|\epsilon_m-\epsilon_n|$,
the operators $\ti{W}^{(\pm)}_{\kappa}$ read
$\ti{W}^{(\pm)}_{\kappa}=\ti{\Gamma}^{(\pm)}_{\kappa}W_{\kappa}$,
where $\ti{\Gamma}^{(\pm)}_{\kappa}$ are defined by
$\ti{\Gamma}^{(+)}_{\kappa}=g(\omega_{\kappa})|\gamma_{\kappa}|^2
(\bar{n}_{\omega_{\kappa}}+1)$ and
$\ti{\Gamma}^{(-)}_{\kappa}=g(\omega_{\kappa})|\gamma_{\kappa}|^2
\bar{n}_{\omega_{\kappa}}$, respectively.
Here $g(\omega)$ is the density of states of the phonon modes,
and $\bar{n}_{\omega}$ is the corresponding phonon occupation number.
Also, we would like to mention that to arrive at \Eq{Rrho-ph}, we have
inserted the device Kohn-Sham Hamiltonian at the initial equilibrium state,
but not the time-dependent one associated with the later evolution,
into the dissipation terms.
This is the well-know approximation in studying dissipative systems
under external-field driving.
This treatment reduces \Eq{Rrho-ph} to the Lindblad form.

To incorporate the effect of electron-phonon interaction into
the Kohn-Sham master equation (\ref{sgm-t}),
we need to recast ${\cal R}_{\rm ph}\rho$ to a RSP density matrix form.
Simple algebra gives
\bea\label{sgm-ph}
&& {\rm Tr}\left\{ a^{\dg}_{\mu}a_{\nu}\sum_{\kappa}
  [W^{\dg}_{\kappa},\ti{W}^{(-)}_{\kappa}\rho
  -\rho\ti{W}^{(+)}_{\kappa} ] \right\}    \nl
&\simeq&  \sigma_{\nu\mu}\sum_n[\ti{\Gamma}^{(-)}_{n\nu}+\ti{\Gamma}^{(+)}_{\mu n}]
     \bar{\sigma}_{nn}
     - \bar{\sigma}_{\nu\mu}\sum_n[\ti{\Gamma}^{(+)}_{n\nu}+\ti{\Gamma}^{(-)}_{\mu n}]
     \sigma_{nn}  .  \nl
\eea
In the derivation of this result, the Wick-type factorization such as
$\la a^{\dg}_{\mu}a_n W_{n\nu} \ra\simeq\sigma_{\nu\mu} \bar{\sigma}_{nn}$
is assumed.
We would like to note that
\Eq{sgm-ph} coincides precisely with the central result derived
by Gebauer and Car in an alternative transport approach \cite{Geb05}.
After simple basis transformation (i.e. from eigenstates to local atomic orbitals ),
inserting \Eq{sgm-ph} into the Kohn-Sham master equation (\ref{sgm-t})
leads to an elegant scheme which can also account for
the electron-phonon scattering in the time-dependent transport process.
We stress that this is another significant advantage of the proposed transport approach.


\section{Conclusion and Discussion}

To summarize, based on the quantum master equation approach,
we have constructed
a practical scheme for the first-principles study
of time-dependent quantum transport.
The basic idea is to combine the transport master-equation with the well-known
time-dependent density functional theory.
By utilizing the partitioning-free initial condition, the scheme is reliably based
on the Runge-Gross theorem. Then, with the help of spectral decomposition,
a closed set of equations for the RSP density matrices
are obtained. Time propagation of this set of equations will directly output
the time-dependent transport current.

It is of interest to note that it is the non-Markovian (but not the Markovian)
master equation that makes us be able to construct the closed set of equations
for the RSP density matrices, based on the spectral decomposition technique.
In Ref.\ \onlinecite{Li05b}, without using the spectral decomposition
and in the framework of Markovian master equation, we established
an alternative form of TDDFT based master equation for the
RSP density matrix. However, in that scheme, the numerical propagation seems
inefficient, because of the non-local effect of time integration.
On the contrary,
the numerical implementation of the present scheme should be efficient
and straightforward. Moreover, it will be very flexible
for the parametrization of highly structured spectral densities,
which makes the description far beyond the Markovian approximation.
Systematic applications and numerical implementations of the proposed
scheme are in progress and will appear in the forthcoming publications.

Finally, we remark that the major approximation involved in the master equation
is the second-order Born approximation for the tunnel Hamiltonian.
This is a standard and well-justified approximation,
which makes the resultant master equation good enough in a large number of
dissipative systems (e.g. in quantum optics).
Also, our recent work clearly showed its satisfactory
accuracy in quantum transport \cite{Li04,Li05a,Li05b,Luo06}.
In this context, we may roughly claim that its accuracy is
at least at the level of sequential transport.
Noticeably, in the first-principles study, other numerical errors will
completely cover up the inaccuracy of sequential transport.
Therefore, the proposed TDDFT master equation scheme should be an attracting
theoretical tool for the first-principles study of time-dependent transport.

\appendix
\section{``$n$"-Resolved Master Equation}

In this appendix we present the derivation of the non-Markovian form of the
``$n$"-resolved master equation.
In almost the same spirit of Ref.\ \onlinecite{Li05a},
let us regard
$H' = \sum_{\mu} \left( a^{\dg}_{\mu} F_{\mu} + \rm{H.c.}\right)$
as a coupling of the system of interest to a dissipative environment.
Treating $H'$ as perturbation and up to the second order,
a formal equation for the reduced density matrix is obtained as \cite{Yan98}
\bea\label{ME-1a}
\dot{\rho}(t)
= -i {\cal L}\rho(t) - \int^{t}_{0}d\tau \la {\cal L}'(t){\cal G}(t,\tau)
                {\cal L}'(\tau) \ra \rho(\tau).
\eea

Here the Liouvillian superoperators are defined as
${\cal L}(\cdots)\equiv [H_S,(\cdots)]$,
${\cal L'}(\cdots)\equiv [H',(\cdots)]$, and
${\cal G}(t,\tau)(\cdots)\equiv G(t,\tau)(\cdots)G^{\dg}(t,\tau)$
with $G(t,\tau)$ the usual propagator (Green's function) associated with
the central system (device) Hamiltonian $H_S$.
The reduced density matrix $\rho(t)=\rm{Tr}_B[\rho_T(t)]$,
and $\la (\cdots)\ra=\rm{Tr}_B[(\cdots)\rho_B]$ with $\rho_B$
the density matrix of the electrode reservoirs.

The trace in \Eq{ME-1a} is over all the electrode degrees of freedom,
leading thus to the equation of motion of the {\it unconditional}
reduced density matrix of the central system.
However, more information is to be contained if we keep track of the record
of electron numbers arrived at the collector (right electrode).
We therefore classify the Hilbert space of the electrodes as follows.
First, we define the subspace in the absence of electron
arrived at the collector as
``$B^{(0)}$", which is spanned by the product
of all many-particle states of the two isolated reservoirs, formally denoted
as $B^{(0)}\equiv\mb{span}\{|\Psi_L\ra\otimes |\Psi_R\ra \}$.
Then, we introduce the Hilbert subspace ``$B^{(n)}$" ( $n=1,2,\cdots$),
corresponding to ``$n$" electrons arrived at the collector.
The entire Hilbert space of the two electrodes is $B=\oplus_n B^{(n)}$.

With the above classification of the reservoir states, the average over states
in the entire Hilbert space ``$B$" in \Eq{ME-1}
is replaced with states in the subspace ``$B^{(n)}$",
leading to a {\it conditional} master equation
\bea\label{ME-2}
\dot{\rho}^{(n)}(t) &=& -i{\cal L}\rho^{(n)}(t) - \int^{t}_{0}d\tau
      \mb{Tr}_{B^{(n)}} [{\cal L}'(t){\cal G}(t,\tau)   \nl
      & &  \times {\cal L}'(\tau) \rho_T(\tau)] .
\eea
Here $\rho^{(n)}(t)=\mb{Tr}_{B^{(n)}}[\rho_T(t)]$,
which is the reduced density matrix of the system
{\it conditioned} by the number of electrons arrived at the collector
until time $t$.
Now we transform the Liouvillian operator product in \Eq{ME-2}
into the conventional Hilbert form:
\bea\label{L-H}
&&  {\cal L}'(t){\cal
G}(t,\tau) {\cal L}'(\tau) \rho_T(\tau) \nl
&=& [ H'(t)G(t,\tau)H'(\tau)\rho_T(\tau)G^{\dg}(t,\tau)  \nl & &
- G(t,\tau)H'(\tau)\rho_T(\tau)G^{\dg}(t,\tau)H'(t)] + \mb{H.c.} \nl
&\equiv& [I-II]+\mb{H.c.}
\eea
To proceed, two physical considerations are further taken into account as follows:
(i)
Instead of the conventional Born approximation for the entire density matrix
$\rho_T(\tau)\simeq\rho(\tau)\otimes\rho_B$,
we propose the ansatz $\rho_T(\tau)\simeq\sum_n\rho^{(n)}(\tau)\otimes\rho_B^{(n)}$,
where $\rho_B^{(n)}$ is the density operator of the electron reservoirs
associated with $n$-electrons arrived at the collector.
With this ansatz for the density operator, tracing over the subspace
``$B^{(n)}$" yields
\begin{subequations}\label{n-ave}
\bea \mb{Tr}_{B^{(n)}}[I]&=& \sum_{\mu,\nu} \left\{
           \mb{Tr}_B [F_{\mu}^{\dg}(t)F_{\nu}(\tau)\rho_B^{(n)}] \right. \nl
       & & \times [a_{\mu}G(t,\tau)a_{\nu}^{\dg}\rho^{(n)}(\tau)G^{\dg}(t,\tau)]
\nl
       & &  +   \mb{Tr}_B [F_{\mu}(t)F_{\nu}^{\dg}(\tau)\rho_B^{(n)}]  \nl
       & & \left. \times [a_{\mu}^{\dg}G(t,\tau)a_{\nu}\rho^{(n)}(\tau)G^{\dg}(t,\tau)]
           \right\} \nl
           \\
\mb{Tr}_{B^{(n)}}[II]&=& \sum_{\mu,\nu} \left\{
           \mb{Tr}_B[f_{L\nu}^{\dg}(\tau)\rho_B^{(n)}f_{L\mu}(t)] \right. \nl
       & & \times [ G(t,\tau)a_{\nu}\rho^{(n)}(\tau)G^{\dg}(t,\tau)a_{\mu}^{\dg} ]   \nl
       & & + \mb{Tr}_B [f_{L\nu}(\tau)\rho_B^{(n)}f_{L\mu}^{\dg}(t)] \nl
       & & \times [ G(t,\tau)a_{\nu}^{\dg}\rho^{(n)}(\tau)G^{\dg}(t,\tau)a_{\mu} ] \nl
       &  &  +\mb{Tr}_B[f_{R\nu}^{\dg}(\tau)\rho_B^{(n-1)}f_{R\mu}(t)] \nl
       & & \times [ G(t,\tau)a_{\nu}\rho^{(n-1)}(\tau)G^{\dg}(t,\tau)a_{\mu}^{\dg} ]   \nl
       & & + \mb{Tr}_B [f_{R\nu}(\tau)\rho_B^{(n+1)}f_{R\mu}^{\dg}(t)] \nl
       & & \left. \times [ G(t,\tau)a_{\nu}^{\dg}
            \rho^{(n+1)}(\tau)G^{\dg}(t,\tau)a_{\mu} ] \right\}. \nl
\eea
\end{subequations}
Here we have utilized the orthogonality between states in different subspaces,
which in fact leads to the term selection from the entire density operator $\rho_T$.
(ii) Due to the closed nature of the transport circuit, the extra
electrons arrived at the collector will flow back into
the emitter (left reservoir) via the external circuit. Also, the rapid
relaxation processes in the reservoirs will quickly bring the
reservoirs to the local thermal equilibrium state determined by
the chemical potentials.
As a consequence, after the procedure
(i.e. the state selection) done in \Eq{n-ave}, the electron reservoir
density matrices $\rho_B^{(n)}$ and $\rho_B^{(n\pm 1)}$
should be replaced by $\rho_B^{(0)}$, i.e., the
local thermal equilibrium reservoir state.
Then, the non-Markovian ``$n$"-resolved master equation \Eq{ME-1}
is obtained.

\section{Spectral Decomposition}

In this appendix we show how to express the correlation functions
$C^{(-)}_{\alpha\mu\nu}(t,t')=\la f_{\alpha\mu}(t)f^{\dg}_{\alpha\nu}(t')\ra$
and $C^{(+)}_{\alpha\nu\mu}(t',t)=\la f^{\dg}_{\alpha\nu}(t')f_{\alpha\mu}(t)\ra$
in terms of the sum of exponential functions,
via the technique of {\it spectral decomposition}.
For constant bias voltage,
let us re-express, for instance, $C^{(+)}_{\alpha\nu\mu}(t',t)$ as
\bea \label{B1}
C_{\alpha\nu\mu}^{(+)}(t',t)
  &=&\sum_k t^*_{\alpha\nu k}t_{\alpha\mu k}
    e^{-i\epsilon_k (t-t')} n_{\alpha}(\epsilon_k)  \nl
  &=& \int \frac{d\epsilon}{2\pi} \Gamma^{\alpha}_{\nu\mu}(\epsilon)
      n_{\alpha}(\epsilon) e^{-i\epsilon (t-t')}  ,
\eea
where $\Gamma^{\alpha}_{\nu\mu}(\epsilon)=2\pi\sum_k
       t^*_{\alpha\nu k}t_{\alpha\mu k}\delta(\ep-\ep_k)$
is the spectral density function of the $\alpha$-th electrode,
which can be obtained by the surface Green's function technique
as described in Sec.\ III(C).

To express $C_{\alpha\nu\mu}^{(+)}(t',t)$ as a sum of exponential functions,
let us parameterize the spectral density as follows
\bea \label{B2}
\Gamma^{\alpha}_{\nu\mu}(\epsilon)
= \sum^m_{k=1} \frac{p^{k}_{\alpha\nu\mu}}{(\ep-\Omega^k_{\alpha\nu\mu})^2
  +(\Gamma^k_{\alpha\nu\mu})^2} .
\eea
By employing the theorem of residues, the integral result of \Eq{B1} reads
\bea \label{B3}
C_{\alpha\nu\mu}^{(+)}(t',t)
&=& \sum^{m}_{k=1} \frac{p^k_{\alpha\nu\mu}}{2\Gamma^k_{\alpha\nu\mu}}
   n_{\alpha}(\ti{\Omega}^k_{\alpha\nu\mu}) e^{-i\ti{\Omega}^k_{\alpha\nu\mu}(t-t')} \nl
&& + \frac{i}{\beta} \sum^{\infty}_{n=0}
   \Gamma^{\alpha}_{\nu\mu}(\ep_n) e^{-i\ep_n (t-t')} ,
\eea
where
$\ti{\Omega}^k_{\alpha\nu\mu}=\ti{\Omega}^k_{\alpha\nu\mu}-i\Gamma^k_{\alpha\nu\mu}$,
and $\ep_n=\ep_F-i(2n+1)/\beta$.
Similarly, the correlation function
$C^{(-)}_{\alpha\mu\nu}(t,t')$ can be decomposed.
Then, these two functions can be formally re-expressed as \Eq{Ctt+-}.

About the spectral decomposition, a few remarks are made as follows:
(i) This spectral decomposition technique has been successfully
applied to the non-Markovian dissipative systems \cite{MT99},
and to the quantum transport through molecular wires very recently \cite{Wel05}.
The most prominent advantage of this technique is its flexibility.
That is, it is extremely well suited
for the parametrization of highly structured spectral
densities, leading to long and oscillatory correlation functions,
thus making the description far beyond the Markovian approximation.
In the context of quantum transport, this decomposition technique
avoids the usual wide-band approximation, and can handle
arbitrary band structures which may be obtained even by the first-principles
calculation.
(ii) In reality, we might find
different sets of parameters which can approximate a given
spectral density to the same degree of accuracy.
Nevertheless, this will not lead to a different dynamical
behavior, since the dynamics is determined by the spectral density itself.
(iii) Not all types of spectral densities need the numerical
decomposition of the spectral densities.
For instance, for the Drude spectral density
$J(\omega)=\eta\omega /[1+(\omega/\omega_d)^2]$,
or other spectral densities with simple poles,
there is no approximation other than the finite
number of Matsubara terms involved.
In practice, the number of fitting terms depends on the shape of spectral density.
For example,
for the Ohmic spectral density with exponential cutoff
$J(\omega)=\eta\omega e^{-\omega/\omega_c}$,
only three terms are accurate enough to parametrize the spectral density.
However, for another spectral density of the form
$J(\omega)=\eta\omega^2/(2\omega_c^3) e^{-\omega/\omega_c}$,
nine terms were necessary for an accurate fit \cite{MT99}.
(iv)
For finite temperatures, the contributions of the high Matsubara
frequencies will be small, so
that for practical purposes the summation over the Matsubara
frequencies can be truncated at some value.
Hence, this
decomposition will be most profitable for high temperatures,
when only a few Matsubara frequencies contribute.
For very low temperatures, we may alternatively parametrize the
{\it combined spectral density}, say,
$\ti{\Gamma}^{\alpha}_{\nu\mu}(\epsilon)
 =\Gamma^{\alpha}_{\nu\mu}(\epsilon)n_{\alpha}(\ep)$,
in terms of the Lorentzian spectral function \Eq{B2}.

\vspace{5ex}
{\it Acknowledgments.}
Support from the National Natural Science Foundation of China
(No.\ 90203014, 60376037, and 60425412),
the Major State Basic Research Project No.\ G001CB3095 of China,
and the Research Grants Council of the Hong Kong Government
is gratefully acknowledged.


\end{document}